\documentclass[iop]{emulateapj}

\usepackage{color}

\received{}
\revised{}
\accepted{}

\slugcomment{To Appear in the Astrophysical Journal}

\shorttitle{Can supernovae sustain molecular cloud turbulence from outside?}
\shorttitle{Supernova driven cloud turbulence}
\shortauthors{Seifried et al.}

\begin{document}

\title{Is molecular cloud turbulence driven by external supernova explosions?}

\email{$^\ast$seifried@ph1.uni-koeln.de}

\author{Daniel Seifried$^\ast$}
\affil{I. Physikalisches Institut, Universit\"at zu K\"oln, Z\"ulpicher Str. 77, 50937 K\"oln, Germany}
\author{Stefanie Walch}
\affil{I. Physikalisches Institut, Universit\"at zu K\"oln, Z\"ulpicher Str. 77, 50937 K\"oln, Germany}
\author{Sebastian Haid}
\affil{I. Physikalisches Institut, Universit\"at zu K\"oln, Z\"ulpicher Str. 77, 50937 K\"oln, Germany}
\author{Philipp Girichidis}
\affil{Leibniz-Institut für Astrophysik Potsdam (AIP), An der Sternwarte 16, 14482 Potsdam, Germany\\
Heidelberg Institute for Theoretical Studies, Schloss-Wolfsbrunnenweg 35, 69118 Heidelberg, Germany}
\author{Thorsten Naab}
\affil{Max-Planck-Institut f\"{u}r Astrophysik, Karl-Schwarzschild-Str. 1, 85741 Garching, Germany}

\begin{abstract}
We present high-resolution ($\sim$ 0.1 pc), hydrodynamical and magnetohydrodynamical simulations to investigate whether the observed level of molecular cloud (MC) turbulence can be generated and maintained by external supernova (SN) explosions. The MCs are formed self-consistently within their large-scale galactic environment following the non-equilibrium formation of H$_2$ and CO including (self-) shielding and important heating and cooling processes. The MCs inherit their initial level of turbulence from the diffuse ISM, where turbulence is injected by SN explosions. However, by systematically exploring the effect of individual SNe going off outside the clouds, we show that at later stages the importance of SN-driven turbulence is decreased significantly. This holds for different MC masses as well as for MCs with and without magnetic fields. The SN impact also decreases rapidly with larger distances. Nearby SNe \mbox{($d$ $\sim$ 25 pc)} boost the turbulent velocity dispersions of the MC by up to 70 per cent (up to a few km s$^{-1}$). For $d$ $>$ 50 pc, however, their impact decreases fast with increasing $d$ and is almost negligible. For all probed distances the gain in velocity dispersion decays rapidly within a few 100 kyr. This is significantly shorter than the average timescale for an MC to be hit by a nearby SN under solar neighbourhood conditions ($\sim$ 2 Myr). Hence, at these conditions SNe are not able to sustain the observed level of MC turbulence. However, in environments with high gas surface densities and SN rates ,like the Central Molecular Zone, observed elevated MC dispersions could be triggered by external SNe.
\end{abstract}


\section{Introduction}

Molecular clouds (MCs) are observed to have large non-thermal line widths \citep[e.g.][]{Larson81,Solomon87} indicating supersonic turbulence. This turbulence is present at early as well as later stages when star formation has already set in \citep[e.g][]{Williams94,Hirota11}. However, supersonic turbulence is expected to decay in a turbulent crossing time \citep[e.g.][]{MacLow98,Stone98}, which is typically shorter (a few Myr) than the cloud lifetime \citep[$\sim$ 10 Myr; see e.g.][]{Dobbs14}. Hence, the observed level of turbulence needs to be driven by either internal stellar feedback \citep[see e.g.][for reviews]{Elmegreen04,MacLow04,Ballesteros07}, local gravitational collapse \citep{Ibanez15,Ibanez17}, gas accretion onto the cloud \citep{Klessen10,Goldbaum11}, or supernovae (SNe) located \textit{outside} the cloud \citep{Padoan16}. 

SNe are considered as one of the main drivers of interstellar turbulence \citep[e.g.][]{Elmegreen04,MacLow04,Walch15b,Naab17} and are involved in the formation of MCs e.g. by sweeping up material \citep{Koyama00}. Although there are observations of the interaction of SN remnants with MCs at infrared and $X$-/$\gamma$-ray wavelengths \citep[e.g.][]{Neufeld07,Hewitt09,Tang14}, it remains unclear whether SNe can maintain the internal velocity dispersion of MCs \textit{after} the cloud has formed. To investigate this, realistic MC formation models \textit{including} their surrounding galactic environment, through which the SN shock has to travel before it hits the clouds, are required. Such studies are numerically demanding and recent works by \citet{Ibanez15,Ibanez17} and \citet{Padoan16} come to different conclusions. \citet{Padoan16} argue that external SNe alone can maintain the observed level of turbulence, whereas \citet{Ibanez15,Ibanez17} suggest that gravity rather than SNe causes the observed turbulent velocity dispersion.

Here, we present a novel study, which systematically explores the effect of SNe at varying distances on self-consistently formed MCs embedded in their galactic environment \citep{Seifried17}. The simulations include a chemical network with associated self-consistent cooling and heating processes in combination with a high spatial resolution of \mbox{0.12 pc}

\section{Numerical methods}
\label{sec:ICs}

The simulations presented here make use of the zoom-in technique presented in \citet{Seifried17} for MC formation within the SILCC project \citep[see][for details]{Walch15,Girichidis16}. The simulations are performed with the adaptive mesh refinement code FLASH 4.3 \citep{Fryxell00,Dubey08} using a magnetohydrodynamics solver which guarantees positive entropy and density \citep{Bouchut07,Waagan09}. We model the chemical evolution of the interstellar medium (ISM) using a simplified chemical network for H$^+$, H, H$_2$, C$^+$, CO, e$^-$, and O \citep{Nelson97,Glover07b,Glover10}, which also follows the thermal evolution of the gas including the most relevant heating and cooling processes. The shielding of the interstellar radiation field \citep[$G_0$ = 1.7;][]{Habing68,Draine78} is calculated according to the surrounding column densities of total gas, H$_2$, and CO via the {\sc TreeCol} algorithm \citep{Clark12b,Walch15,Wunsch17}. We solve the Poisson equation for self-gravity with a tree-based method \citep{Wunsch17} and include a background potential from the pre-existing stellar component in the galactic disc, modeled as an isothermal sheet with \mbox{$\Sigma_\mathrm{star}$ = 30 M$_{\sun}$ pc$^{-2}$} and a scale height of \mbox{100 pc.}

Our setup represents a small section of a galactic disc with solar neighborhood properties and a size of 500 pc $\times$ 500 pc $\times$ $\pm$ 5 kpc. The gas surface density is \mbox{$\Sigma_\mathrm{gas}$ = 10 M$_{\sun}$ pc$^{-2}$} and the initial vertical distribution of the gas has a Gaussian profile with a scale height of 30 pc and a midplane density of 9 $\times$ $10^{-24}$ g cm$^{-3}$.
The gas near the disc midplane has an initial temperature of 4500 K and consists of atomic hydrogen and C$^+$.

Up to \mbox{$t_0$ = 11.9 Myr}, we drive turbulence in the disc with SNe, which are randomly placed in the midplane and have a Gaussian distribution with a scale height of 50 pc in the vertical direction. The SN rate is constant at 15 SNe Myr$^{-1}$, corresponding to the Kennicutt-Schmidt star formation rate surface density for \mbox{$\Sigma_\mathrm{gas}$ = 10 M$_{\sun}$ pc$^{-2}$} \citep{Kennicutt98}. For a single SN we inject 10$^{51}$ erg in the form of thermal energy if the Sedov-Taylor radius is resolved with at least 4 cells. Otherwise, we heat the gas within the injection region to \mbox{$10^4$ K} and inject the momentum, which the swept-up shell has gained at the end of the pressure-driven snowplow phase \citep[see][and \citealt{Gatto15} for details]{Blondin98}.

The base grid resolution is 3.9 pc up to $t_0$. At $t_0$ we stop further SN explosions. We choose two regions in which MCs -- henceforth denoted as MC1 and \mbox{MC2 --} are about to form and we continue the simulation for another 1.5 Myr over which we progressively increase the spatial resolution d$x$ in these two regions from 3.9 pc to 0.12 pc \citep[][Table~2]{Seifried17}.

In order to estimate the time required to develop a realistic three-phase ISM from the quiescent initial conditions, we assume a mean expansion speed of our SN remnants of \mbox{$\sim$ 20 -- 30 km s$^{-1}$} \mbox{($\simeq$ 20 -- 30 pc Myr$^{-1}$)} and a mean SN age of \mbox{$t_0$/2 = 6 Myr}. Hence, each SN remnant would cover on average a spherical region with about 240 -- 360 pc in diameter. Given the SN rate of \mbox{15 SNe Myr$^{-1}$}, \mbox{180 SNe} are inserted up to $t_0$. Comparing the area which would be covered by all 180 SNe to the area of the simulated part of the disk (500 pc $\times$ 500 pc) gives us the typical number of how often each point in the disk is hit by a SN shock. With the numbers above we find that up to $t_0$ each point in the disk midplane is affected on average by $\sim$ 50 SNe. Although this is a rather crude estimate, it shows that a time of $\sim$ 12 Myr is sufficient to develop a self-consistent, turbulent three-phase ISM.

\begin{figure}
 \includegraphics[width=\linewidth]{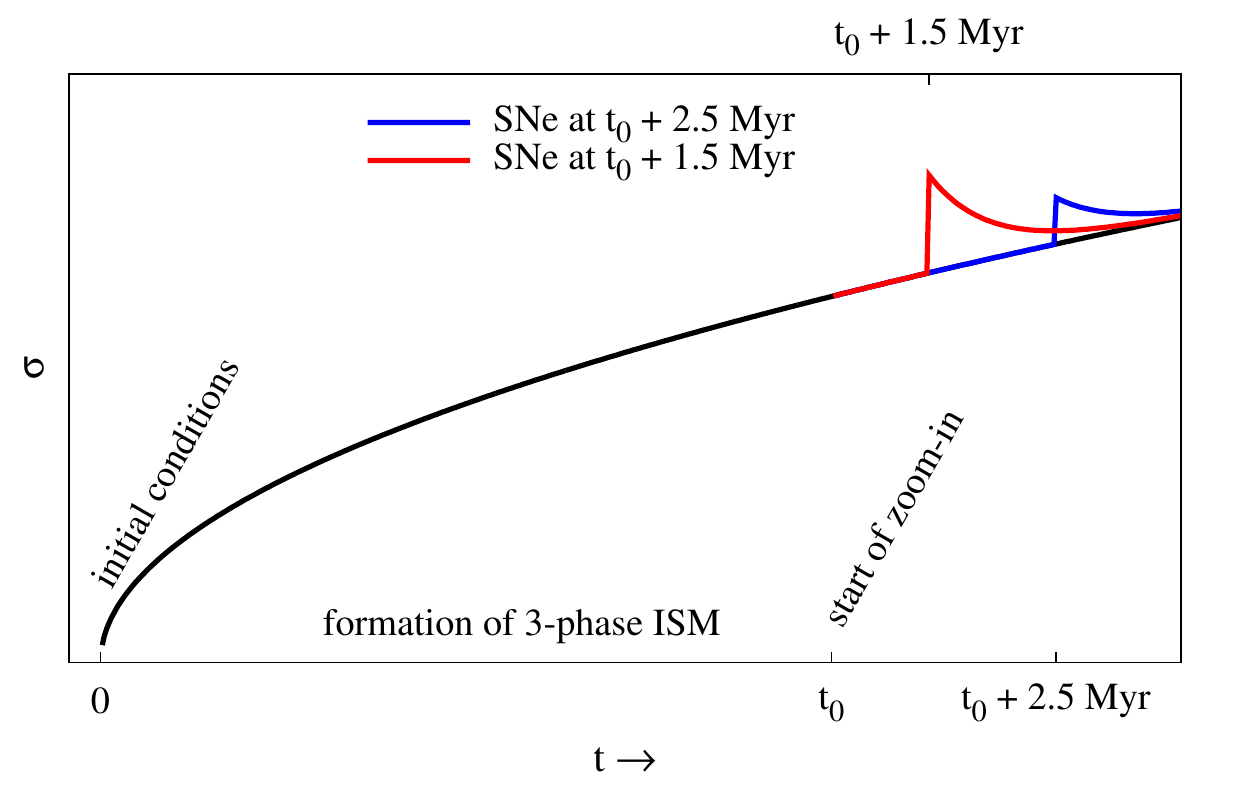}
 \caption{Schematic view of the different phases in the simulations showing the initial phase up to $t_0$ for the evolution of the three-phase medium and the different times $t_\mathrm{SN}$ =  $t_0$ + 1.5 Myr and $t_0$ + 2.5 Myr at which SNe are inserted for the parameter study. For the runs with magnetic fields (Section~\ref{sec:MHD}), we insert SNe at $t$ = $t_\mathrm{0, \, mag}$ + 1.5 and $t_\mathrm{0, \, mag}$ + 2.5 Myr.}
 \label{fig:overview}
\end{figure}

Shortly after the clouds are fully refined, at \mbox{$t_\mathrm{SN}$ = $t_0$ + 1.53 Myr} we explode single SNe at different distances of $d$ = 25, 50, 62.5, or 75 pc from the center-of-mass of either cloud along the $\pm x$-, $\pm y$-, or $\pm z$-direction. In addition, we also investigate the impact of SNe going off $\sim$ 1 Myr later at \mbox{$t_\mathrm{SN}$ = $t_0$ + 2.5 Myr} (Section~\ref{sec:laterSN}). A schematic overview of the different times in our simulations is shown in Figure~\ref{fig:overview}.

Furthermore, in Section~\ref{sec:MHD} we investigate the influence of magnetic fields on our results. For this, we zoom-in on 2 clouds denoted as MC3 and MC4 at $t_\mathrm{0, \, mag}$ = 16 Myr in a comparable simulation with magnetic fields. In total, we consider 132 different simulations including SNe  (see Table~\ref{tab:overview}), thus giving us statistically reliable results.

\begin{table}
 \caption{Overview of the performed simulations showing the cloud name, the time $t_0$, at which we start to zoom-in, the distance $d$ and time $t_\mathrm{SN}$ the SN goes off, and whether magnetic fields are included.}
 \centering
 \begin{tabular}{ccccc}
  \hline
  cloud & $d$ [pc] & $t_0$ [Myr] &$t_\mathrm{SN}$ [Myr] & B-field \\
  \hline
  MC1 & 25, 50, 62.5, 75 & 11.9 & 13.4 & No \\
  MC2 & 25, 50, 62.5, 75 & 11.9 & 13.4 & No \\
  MC1 & 25, 50 & 11.9 & 14.4 & No \\
  MC2 & 25, 50 & 11.9 & 14.4 & No \\
  \hline
  MC3 & 25, 50, 62.5 & 16.0 & 17.5 & Yes \\
  MC4 & 25, 50, 62.5 & 16.0 & 17.5 & Yes \\
  MC3 & 25, 50 & 16.0 & 18.5 & Yes \\
  MC4 & 25, 50 & 16.0 & 18.5 & Yes \\
  \hline
  \vspace{0.2cm}
 \end{tabular}
 \label{tab:overview}
\end{table}

\section{Results}

\subsection{Column density evolution}

\begin{figure*}
 \includegraphics[width=\linewidth]{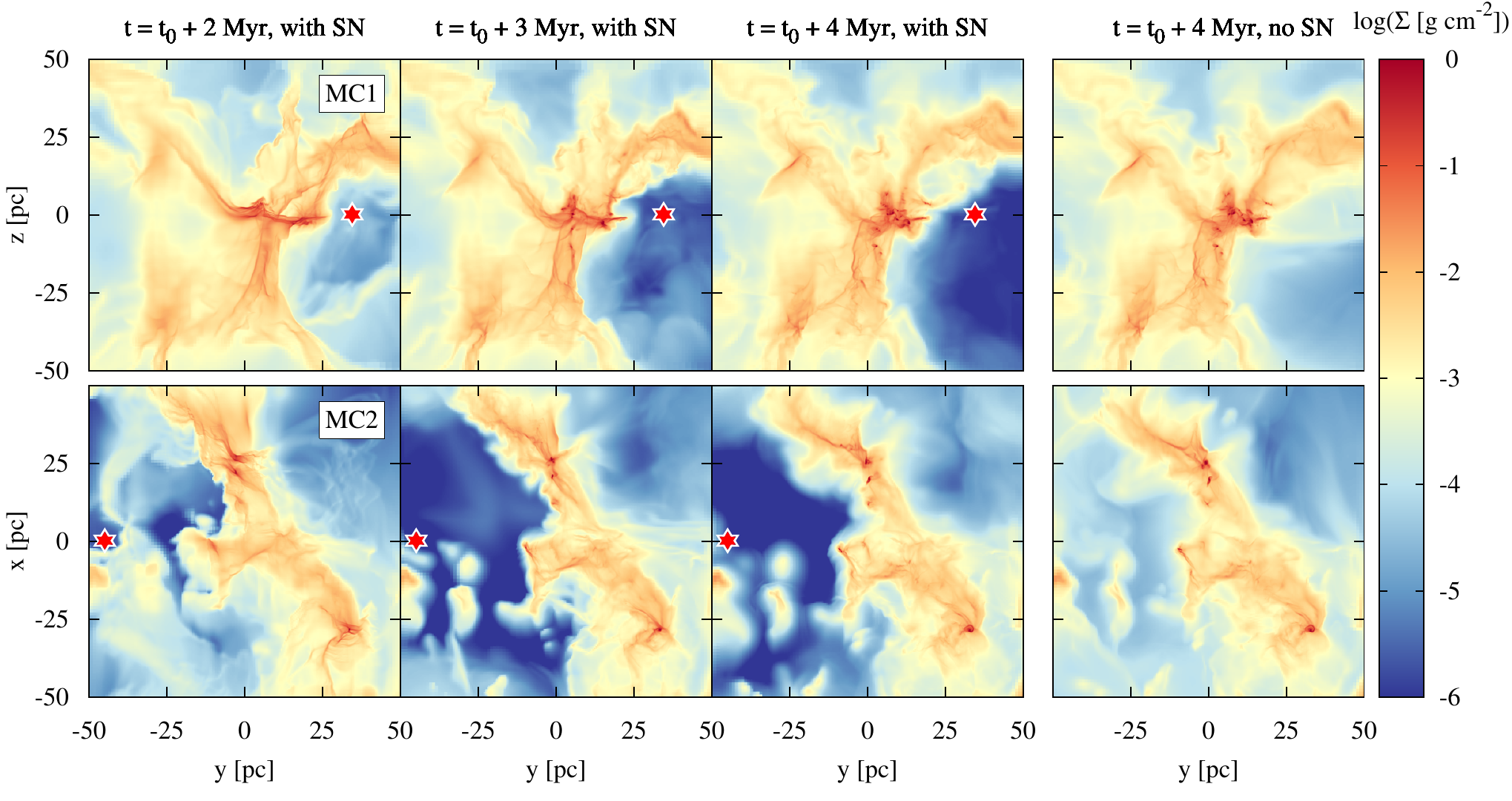}
 \caption{Time evolution ($t$ = \mbox{$t_0$ + 2 Myr} to $t_0$ + 4 Myr) of the column density of MC1 (top row) and MC2 (bottom row) as they are hit by a nearby SN (red star). For comparison the right panels show the reference run with no SN at $t_0$ + 4 Myr. The comparable structure of the clouds with and without a SN indicates that the SN mainly affects the low-density environment.}
 \label{fig:col_dens}
\end{figure*}

In \citet{Seifried17} we have investigated the formation and early evolution of MC1 and MC2 for 5 Myr, i.e. up to \mbox{$t_0$ + 5 Myr}, without any further SNe in the simulation domain. These simulations serve as the reference simulations indicated with the subscript ``no SN''. The clouds harbor increasing amounts of molecular gas and develop a strongly fragmented sub-structure (right panels of Figure~\ref{fig:col_dens}).

The left panels show the time evolution of the column density of MC1 (top row) and MC2 (bottom row) which are exposed to a SN at a distance of 25 and 50 pc, respectively. The SN bubbles expand asymmetrically and seem not to be able to significantly affect the central, high-density regions. A qualitatively similar behavior is found for all other simulations.

\subsection{Time evolution}

\begin{figure}
 \includegraphics[width=\linewidth]{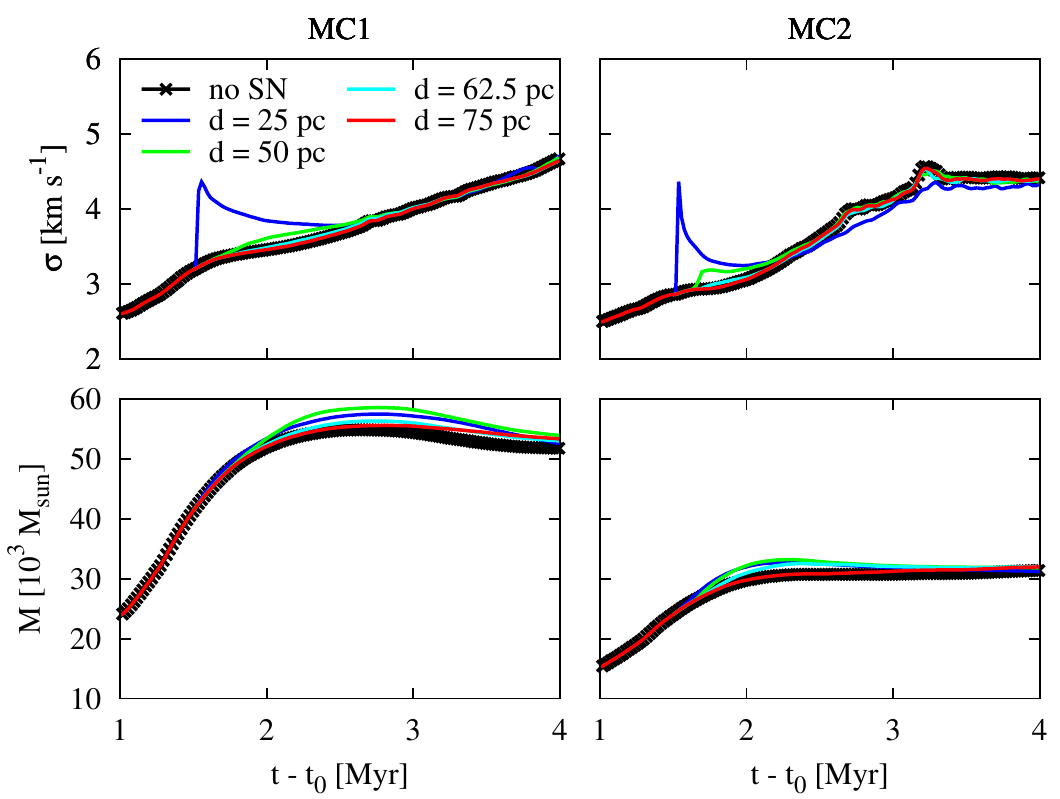}
 \caption{Time evolution of the velocity dispersion (top row) and the mass (bottom row) of the non-magnetic clouds MC1 (left) and MC2 (right) calculated for gas with densities $\geq$ 100 cm$^{-3}$. The black line shows $\sigma_\mathrm{no \, SN}$ and $M_\mathrm{no \, SN}$ from the reference runs, the remaining lines show $\sigma_\mathrm{SN}$ and $M_\mathrm{SN}$ for SNe exploding at distances of 25, 50, 62.5, and 75 pc along the -$y$-direction.}
 \label{fig:sigma_mass}
\end{figure}
\begin{figure}
 \includegraphics[width=\linewidth]{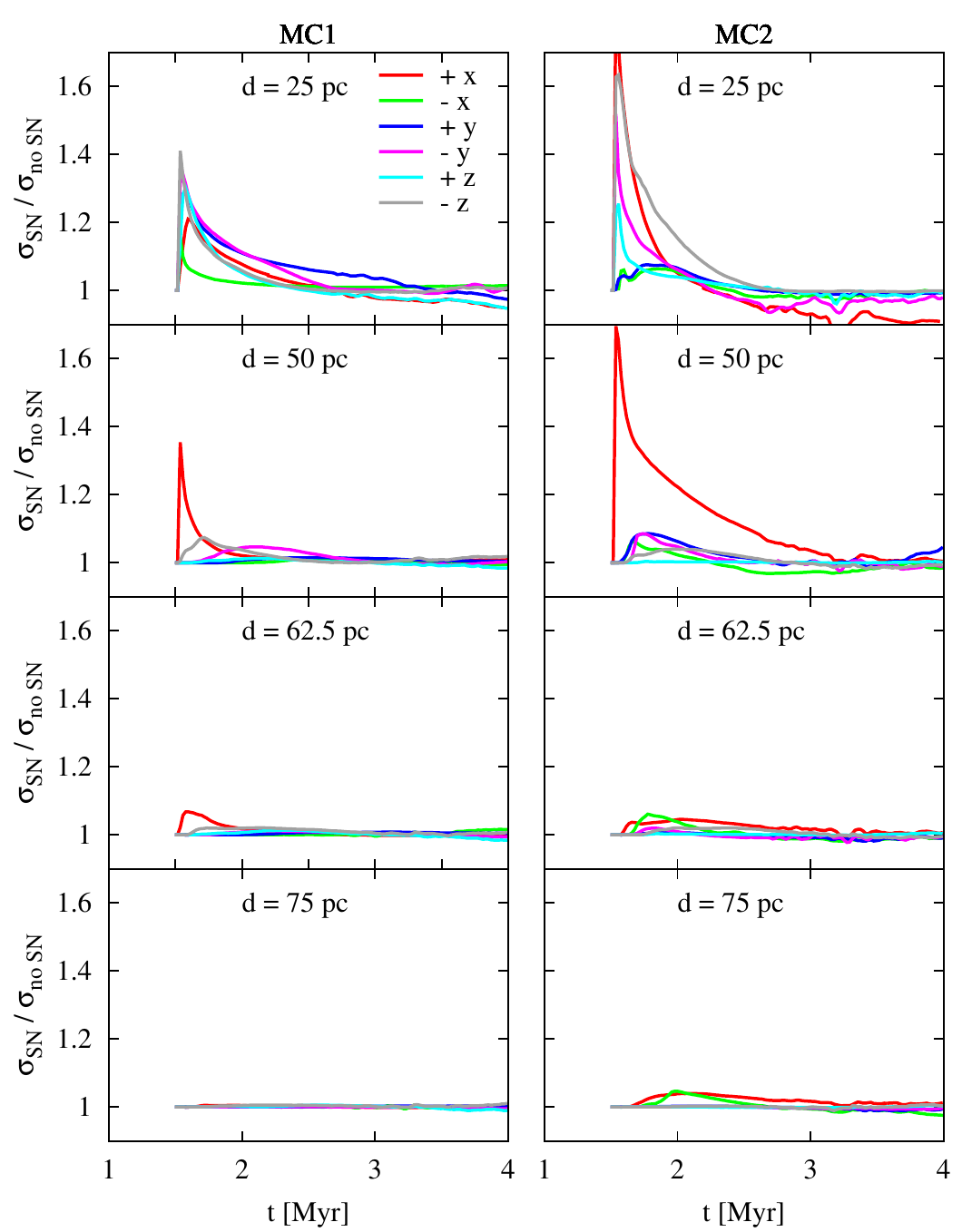}
 \caption{Relative change, $\sigma_\mathrm{SN}$/$\sigma_\mathrm{no \, SN}$, for SNe going off at increasing distances (from top to bottom) along the $\pm x$-, $\pm y$-, and $\pm z$-directions for  MC1 (left) and MC2 (right) calculated for gas with densities $\geq$ 100 cm$^{-3}$. At distances up to 50 pc, $\sigma$ is temporarily (for $\sim 100$ kyr) enhanced up to 70\% compared with the respective reference simulation. At larger distances the SNe have very little effect. }
 \label{fig:sigma_rel}
\end{figure}

In Figure~\ref{fig:sigma_mass} we show the time evolution of the mass and velocity dispersion of the two non-magnetic clouds. We define the mass $M$ of the cloud as the mass of all gas with particle densities \mbox{$n$ $\geq$ 100 cm$^{-3}$} (assuming a mean molecular weight of $\mu$ = 2.3). The velocity dispersion $\sigma$ of this mass is then defined as the mass-weighted root-mean-square velocity with respect to the center-of-mass (COM) velocity of the cloud, $\vec{v}_\mathrm{COM}$, i.e.
\begin{equation}
 \sigma = \sqrt{ \frac{ \Sigma \, ( \mathrm{d}m \times (\vec{v} - \vec{v}_\mathrm{COM})^2)}{\Sigma \, \mathrm{d}m} } \, ,
\end{equation}
where d$m$ and $\vec{v}$ are the mass and velocity of an individual cell, respectively, and we sum over all cells with \mbox{$n$ $\geq$ 100 cm$^{-3}$}.

For the unperturbed clouds, the masses $M_\mathrm{no \, SN}$ increase over about 2 Myr and afterward remain roughly constant, with values around 5 $\times$ 10$^4$ M$_{\sun}$ and 3 $\times$ 10$^4$ M$_{\sun}$ for MC1 and MC2, respectively (bottom row of Figure~\ref{fig:sigma_mass}, black line with crosses). The velocity dispersion of the unperturbed clouds $\sigma_\mathrm{no \, SN}$ (top row) increases continuously for MC1, whereas for MC2 around \mbox{$t$ $\simeq$ $t_0$ + 3 Myr} it levels off at \mbox{$\sigma_\mathrm{no \, SN}$ $\simeq$ 4.5 km s$^{-1}$}. We emphasize that both, the derived masses and velocity dispersions indicate that the chosen clouds are representative of typical MCs in the Milky Way, for example \citet{Miville17} found a median cloud mass of \mbox{4 $\times$ 10$^4$ M$_\sun$} \citep[but see also e.g.][]{Solomon87,Elmegreen96,Heyer01,Roman10}. Furthermore, the high level of turbulence of $\sim$ 2.5 km s$^{-1}$ present already at the beginning of the cloud evolution as well as the strongly fragmented sub-structure (Figure~\ref{fig:col_dens}) indicate that the initial SN-driving period of 11.9 Myr was sufficient to establish a turbulent three-phase medium.

Considering the impact of external SNe, we find that the cloud masses $M_\mathrm{SN}$ (bottom row of Figure~\ref{fig:sigma_mass}) show only a moderate initial increase by $\sim$ 1\% ($\lesssim$ 1000 $M_\sun$) compared to $M_\mathrm{no \, SN}$, which we attribute to the compression of less dense gas by the incoming SN blast wave. We note that in a few runs, at later times, $M_\mathrm{SN}$ even slightly drops below $M_\mathrm{no \, SN}$, i.e. the SN has a negative effect on the cloud mass, but the changes are always of the order of a few percent. The influence of the SNe on $\sigma$ (top row of Figure~\ref{fig:sigma_mass}), however, is much more pronounced than on $M$. We find that in particular nearby SNe seem to have a noticeable effect on $\sigma$ increasing it by a few km s$^{-1}$. For more distant ($d$ $>$ 50 pc) SNe, the impact is significantly lower.

In Figure~\ref{fig:sigma_rel} we show the relative change of $\sigma_\mathrm{SN}$ with respect to $\sigma_\mathrm{no \, SN}$, now for all directions and the four distances considered in this work. Overall, the SN impact temporarily enhances $\sigma$ by at most 70\% for the nearby SNe ($d$ = 25 and 50 pc). For more distant SNe, the relative increase is almost negligible with $\sim$ 1\% for $d$ = 75 pc. Interestingly, even for nearby explosions, the increase is short-lived and the additional dispersion decays on timescales of a few \mbox{100 kyr}.

\subsection{Maximum gain in velocity dispersion and decay time}
\label{sec:gain}

\begin{figure}
 \includegraphics[width=0.48\textwidth]{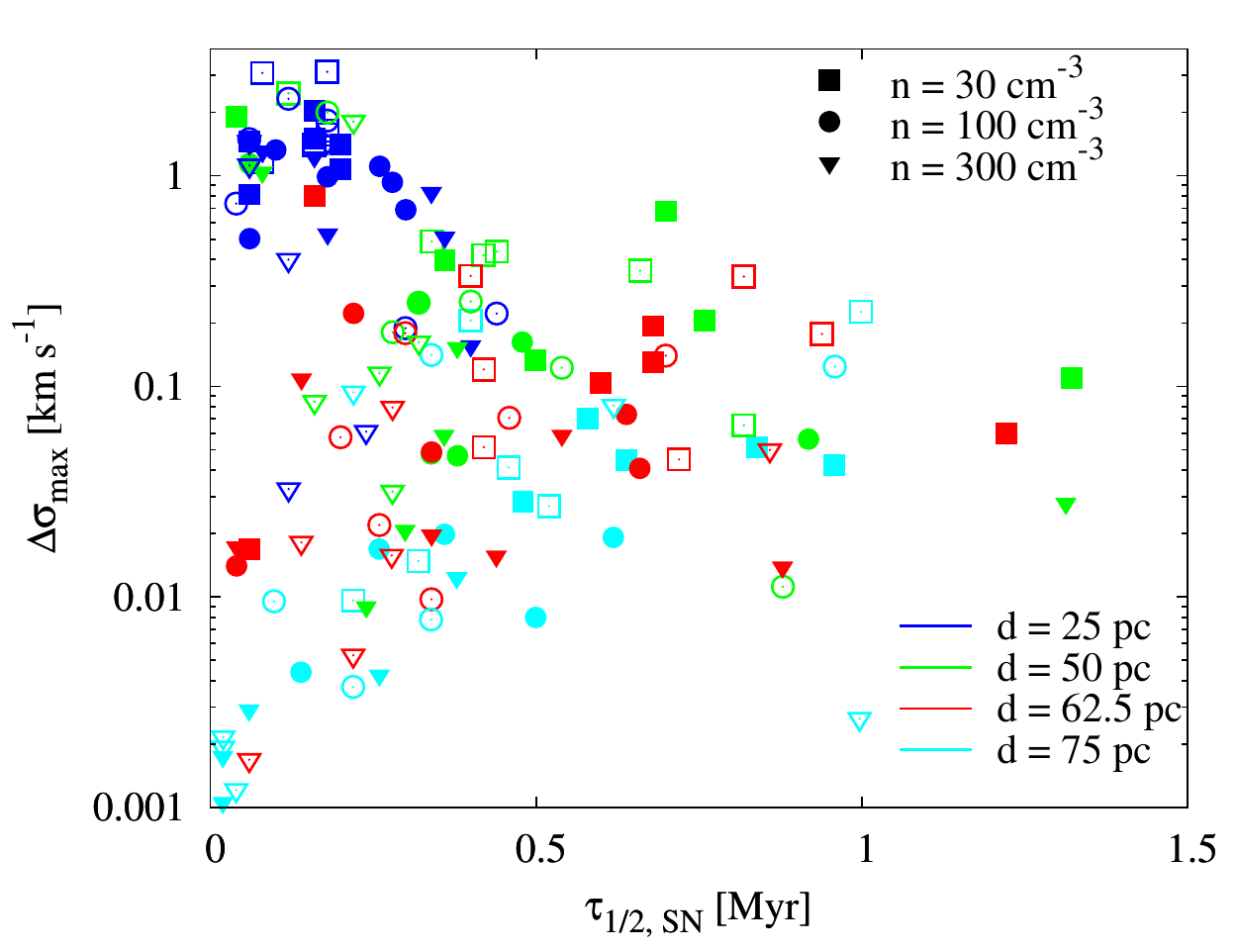}
 \caption{Relation between the decay time $\tau_\mathrm{1/2, SN}$ and the maximum increase $\Delta \sigma_\mathrm{max}$ of the additional turbulent velocity component caused by SNe with different distances (symbols) for MC1 and MC2 (open and closed symbols) calculated using different density thresholds (color-coded).}
 \label{fig:decay_time}
\end{figure}

We determine the additional turbulent velocity dispersion driven by a nearby SN,
\begin{equation}
\mbox{$\Delta \sigma_\mathrm{max}$ =  max($\sigma_\mathrm{SN} - \sigma_\mathrm{no \, SN}$)}
\label{eq:sigma_max}
\end{equation}
reached at $t$ = $t_\mathrm{max}$, and its decay time (half-life), $\tau_\mathrm{1/2, SN}$, which is the time which passes until \mbox{$\sigma_\mathrm{SN} - \sigma_\mathrm{no \, SN}$} drops from $\Delta \sigma_\mathrm{max}$ to \mbox{$0.5 \times \Delta \sigma_\mathrm{max}$}\footnote{Note that the half-life is related to the e-folding time $\tau_\mathrm{e}$ of an exponential decay as $\tau_\mathrm{1/2, SN}$ = $\tau_\mathrm{e}$ $\times$ ln(2). When fitting an exponential function to the decaying part of $\sigma_\mathrm{SN} - \sigma_\mathrm{no \, SN}$, we obtain very similar timescales. However, determining $\tau_\mathrm{1/2, SN}$ provides somewhat more robust estimates.}
, i.e.
\begin{equation}
 (\sigma_\mathrm{SN} - \sigma_\mathrm{no \, SN})(t_\mathrm{max} + \tau_\mathrm{1/2, SN}) = 0.5  \times \Delta \sigma_\mathrm{max} \, .
 \label{eq:tau}
\end{equation}
We do this for both MCs, but now using three different density thresholds of 30, 100, and \mbox{300 cm$^{-3}$}. Considering Figure~\ref{fig:decay_time}, we find that with increasing distance, $\Delta \sigma_\mathrm{max}$ decreases. Furthermore, for \mbox{$d$ $\leq$ 50 pc} there seems to be a weak anti-correlation between $\tau_\mathrm{1/2, SN}$ and $\Delta \sigma_\mathrm{max}$ (and thus a positive correlation between $\tau_\mathrm{1/2, SN}$ and $d$). We attribute this to the fact that turbulence decays roughly within one crossing time, which is inversely proportional to $\sigma$ in the supersonic case \citep{Stone98}. Furthermore, typically $\tau_\mathrm{1/2, SN} \sim 100 - 600$ kyr, i.e. the effect of SNe is very limited in time compared to MC formation and evolution times of up to a few 10 Myr \citep[see e.g. the review by][]{Dobbs14}. Since the SN energy is lost quickly by radiative cooling in the dense gas of the cloud \citep[a few 1000 yr, see e.g][]{Koyama00,Haid16}, $\sigma$ is only temporarily boosted and decays subsequently.

\begin{figure}
 \includegraphics[width=\linewidth]{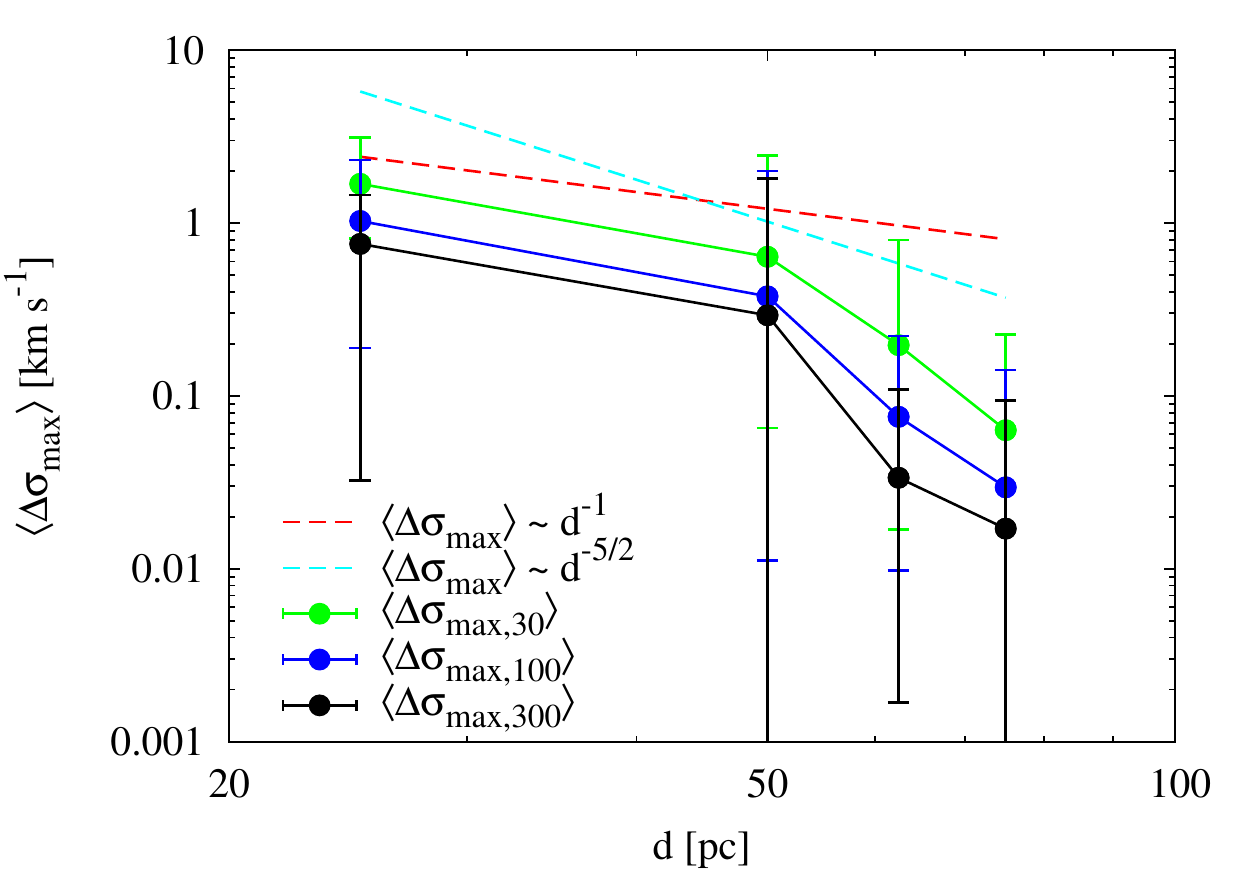}
 \caption{Mean maximum increase of the velocity dispersion $\left\langle \Delta \sigma_\mathrm{max} \right\rangle$ caused by SNe as a function of their distance averaged over MC1 and MC2. The error bar gives the minimum and maximum values found. For $d$ $>$ 50 pc, the velocity gain decreases steeper than expected for adiabatic expansion ($d^{-1}$, red dashed line). For the momentum-conserving snowplow phase we would expect $\left\langle \Delta \sigma_\mathrm{max} \right\rangle \propto d^{-2.5}$ (cyan dashed line).}
 \label{fig:sigma_distance}
\end{figure}
Combining the results of Figure~\ref{fig:decay_time} for MC1 and MC2, we calculate the mean value of $\Delta \sigma_\mathrm{max}$ for each $d$ (Figure~\ref{fig:sigma_distance}). The error bars indicate the minimum and maximum value found. First, we see that $\left\langle \Delta \sigma_\mathrm{max} \right\rangle$ decreases with increasing density threshold. This is due to the dissipation of kinetic energy as the shock travels deeper into the cloud, i.e. into higher-density regions. Furthermore, for \mbox{$d$ $\leq$ 50 pc},  $\left\langle \Delta \sigma_\mathrm{max} \right\rangle$ is typically between a few 0.1 - 1 km s$^{-1}$, whereas for \mbox{$d$ = 62.5 pc}, \mbox{$\left\langle \Delta \sigma_\mathrm{max} \right\rangle$ $\simeq$ 0.1 km s$^{-1}$}, i.e. less than the typical sound speed $c_\mathrm{s} \simeq$ 0.2 km s$^{-1}$ of molecular gas with \mbox{$T$ = 10 K}. For \mbox{$d$ = 75 pc}, $\left\langle \Delta \sigma_\mathrm{max} \right\rangle$ only increases by a few 10 m s$^{-1}$.

Assuming energy conservation, which is applicable if the SN is hitting the cloud during the Sedov-Taylor stage, we obtain the following for the energy gain of the cloud:
\begin{equation}
 \frac{1}{2} M_\mathrm{MC} \Delta \sigma_\mathrm{max}^2 = \frac{A_\mathrm{MC}}{4 \pi d^2} \times E_\mathrm{SN} \times 0.27 \, .
 \label{eq:energy}
\end{equation}
In the following we assume fiducial values of \mbox{$A_\mathrm{MC}$ = $\pi \times$ (10 pc)$^2$} and \mbox{$M_\mathrm{MC} = 5 \times 10^4$ M$_{\sun}$} for the area and mass of the MC, and take into account that only 27\% of the SN energy ($E_\mathrm{SN}$ = 10$^{51}$ erg) is in kinetic form during the Sedov-Taylor stage. This then results in a relation \mbox{$\left\langle \Delta \sigma_\mathrm{max} \right\rangle$ $\propto$ $d^{-1}$} (red dashed line in Figure~\ref{fig:sigma_distance}), which is similar to the observed slope for $d$ $\le$ 50 pc.

Furthermore, for the typical number densities of 0.2 -- 5 cm$^{-3}$ of the regions, within which the SNe explode, the Sedov-Taylor radii are $r_\mathrm{ST}$ $\simeq$ 5 -- 35 pc \citep{Haid16} and are always smaller than the distances of the SNe. Hence, the SN must have lost a significant amount of energy due to radiative cooling even \textit{before} hitting the cloud. In particular for $d$ $>$ 50 pc, where the SN has entered the momentum-conserving snowplow phase, the decline of $\left\langle \Delta \sigma_\mathrm{max} \right\rangle$ is much steeper than $d^{-1}$. In this stage we expect an energy gain of the cloud of
\begin{equation}
 \frac{1}{2} M_\mathrm{MC} \Delta \sigma_\mathrm{max}^2 = \frac{A_\mathrm{MC}}{4 \pi d^2} \times \frac{p_\mathrm{shell}^2}{2 M_\mathrm{shell}} \, ,
 \label{eq:momentum}
\end{equation}
assuming that all the mass in the volume occupied by the SN remnant is contained in the swept-up shell, i.e. $M_\mathrm{shell} \propto d^3$, and that this swept-up shell has gained a momentum $p_\mathrm{shell}$.
This results in \mbox{$\left\langle \Delta \sigma_\mathrm{max} \right\rangle$ $\propto$ $d^{-2.5}$} (cyan dashed line in Figure~\ref{fig:sigma_distance}), which is closer to the decrease of $\left\langle \Delta \sigma_\mathrm{max} \right\rangle$ for $d>50$ pc.
However, at any $d$, the actual values for $\left\langle \Delta \sigma_\mathrm{max} \right\rangle$ are smaller than those of the two theoretical predictions, which indicates that the energy transfer from the SN shock to the MC is not very efficient \citep{Walch15b}.

It is difficult to find a common origin for the scatter of $\Delta \sigma_\mathrm{max}$ at a given distance. We do not see a correlation of $\Delta \sigma_\mathrm{max}$ with the density of the region in which the SN explodes, nor with the integrated column density along the path of the SN shock towards the center of the MC. It seems that $\Delta \sigma_\mathrm{max}$ is highly sensitive to the density distribution along the path of the shock \citep{Haid16} and can be significantly affected even by small fluctuations. This is also supported by the fact that the standard deviation of $\left\langle \Delta \sigma_\mathrm{max} \right\rangle$ increases with increasing distance, i.e. as there are more density fluctuations along the path of the SN shock. Moreover, differences in the cloud structure when seen from the 6 different directions also cause a scatter in $\Delta \sigma_\mathrm{max}$.

\subsection{The influence of magnetic fields}
\label{sec:MHD}

In order to test whether magnetic fields measurably alter our results, we perform a corresponding simulation including magnetic fields. For this we use the same initial conditions as described in Section~\ref{sec:ICs}, but now include a magnetic field along the $x$-direction with a strength of 
\begin{equation}
 B_{x} = B_{x,0} \sqrt{\rho(z)/\rho_0} \, ,
\end{equation}
with \mbox{$B_{x,0}$ = 3 $\mu$G} in accordance with recent observations \citep[e.g.][]{Beck13}. We insert SNe up to \mbox{$t_\mathrm{0, \, mag}$ = 16 Myr} in order to establish a realistic multiphase ISM. We note that we here apply the SN driving for a longer time than for MC1 and MC2 since the magnetic field delays the formation of overdense structures, which subsequently form MCs \citep[][Girichidis et al., submitted]{Walch15,Girichidis16}. Afterwards, we follow the evolution of 2 MCs -- denoted as MC3 and MC4 -- in selected regions with the same zoom-in strategy as before and insert a nearby SN\footnote{Due to the small influence of SNe at $d$ = 75 pc, we do not follow this case further here.} (\mbox{$d$ = 25,} 50, or 62.5 pc) along 6 different directions at \mbox{$t$ = $t_\mathrm{0, \, mag}$ + 1.5 Myr}, i.e. once we have reached the final resolution of 0.12 pc.

\begin{figure}
 \includegraphics[width=\linewidth]{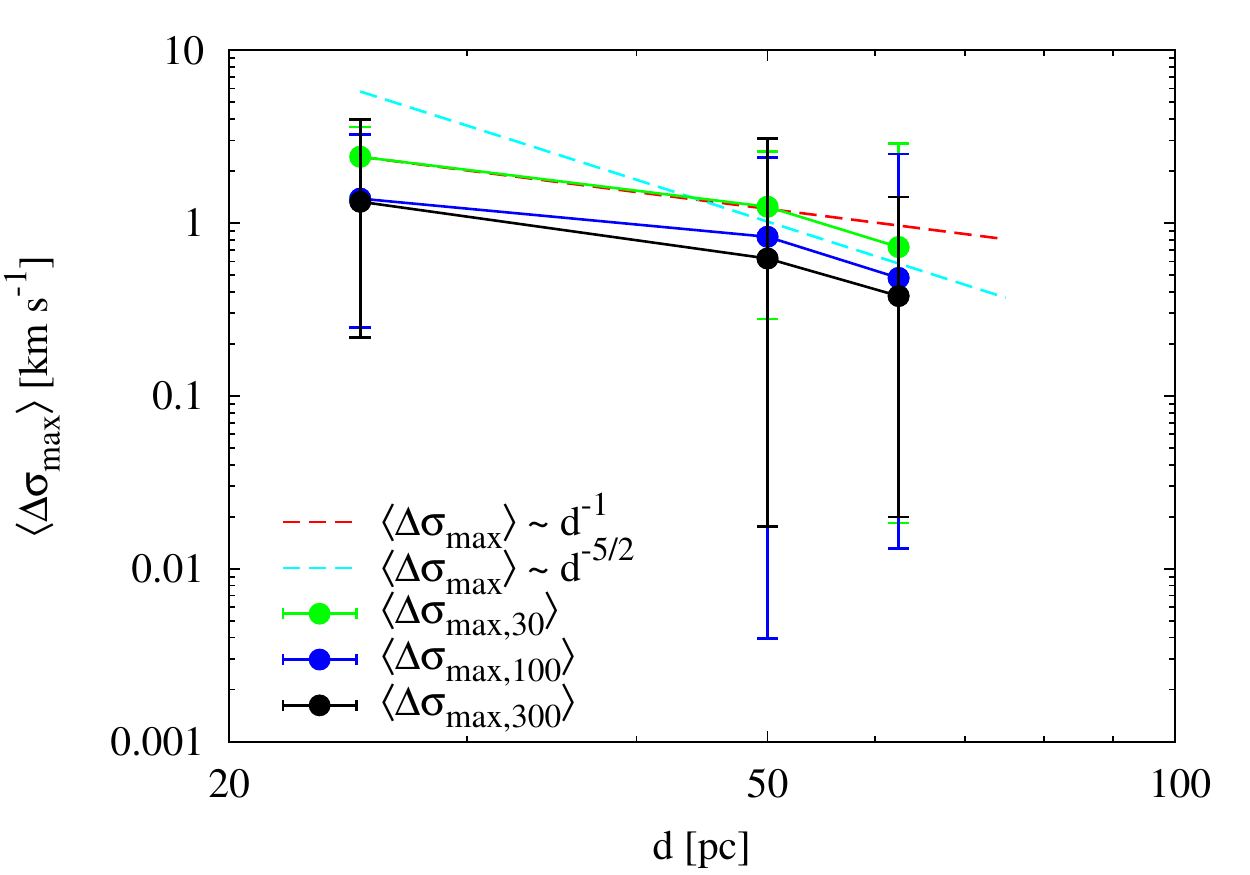}
 \caption{Same as in Figure~\ref{fig:sigma_distance}, but now for the runs MC3 and MC4 including magnetic fields. Overall we see a similar behaviour as for the runs without magnetic fields, in particular the drop of $\left\langle \Delta \sigma_\mathrm{max} \right\rangle$ above \mbox{$d$ = 50 pc.}}
 \label{fig:sigma_distance_mag}
\end{figure}

We redo the same analysis as before by comparing the runs to a reference run without any nearby SN. A short overview of the results of these runs is given in the Appendix. Here, we only focus on the distance dependence of $\Delta \sigma_\mathrm{max}$ in Figure~\ref{fig:sigma_distance_mag}. Overall, we see a similar behaviour as for the runs without magnetic fields (Figure~\ref{fig:sigma_distance}). The gain in $\sigma$ decreases with increasing $d$, in particular at \mbox{$d$ $>$ 50 pc} the decrease becomes steeper. However, now it is closer to the theoretical expectation of a SN in the momentum-conserving snowplow phase than for the unmagnetized runs. For \mbox{$d$ $\leq$ 50 pc}, $\Delta \sigma_\mathrm{max}$ ranges from several \mbox{0.1 km s$^{-1}$} and  \mbox{$\sim$ 3 km s$^{-1}$}, whereas for \mbox{$d$ = 62.5 pc} it drops below \mbox{0.7 km s$^{-1}$}.

The induced level of turbulence is thus somewhat higher compared to the non-magnetic runs (a factor of \mbox{$\sim$ 2}). On the one hand, this could indicate that magnetic fields support the transport of turbulence from the diffuse into the dense ISM. On the other hand, as we will show in Section~\ref{sec:laterSN}, this can partly be attributed to the fact that we model the SN impact at a stage where the clouds are in a relatively early assembly phase with masses of $\sim$ 1.5 $\times$ 10$^4$ M$_\sun$ only (bottom left panel of Fig.~\ref{fig:sigma_mass_mag}), i.e. significantly less massive than the MCs without magnetic fields at this stage (4.1 and 2.4 $\times$ 10$^4$ M$_\sun$ for MC1 and MC2, respectively). However, overall our main findings of a moderate influence of external SNe on MCs seem to also hold in the presence of magnetic fields.

\subsection{SN impact at a later stage}
\label{sec:laterSN}

\begin{figure*}
 \includegraphics[width=0.48\linewidth]{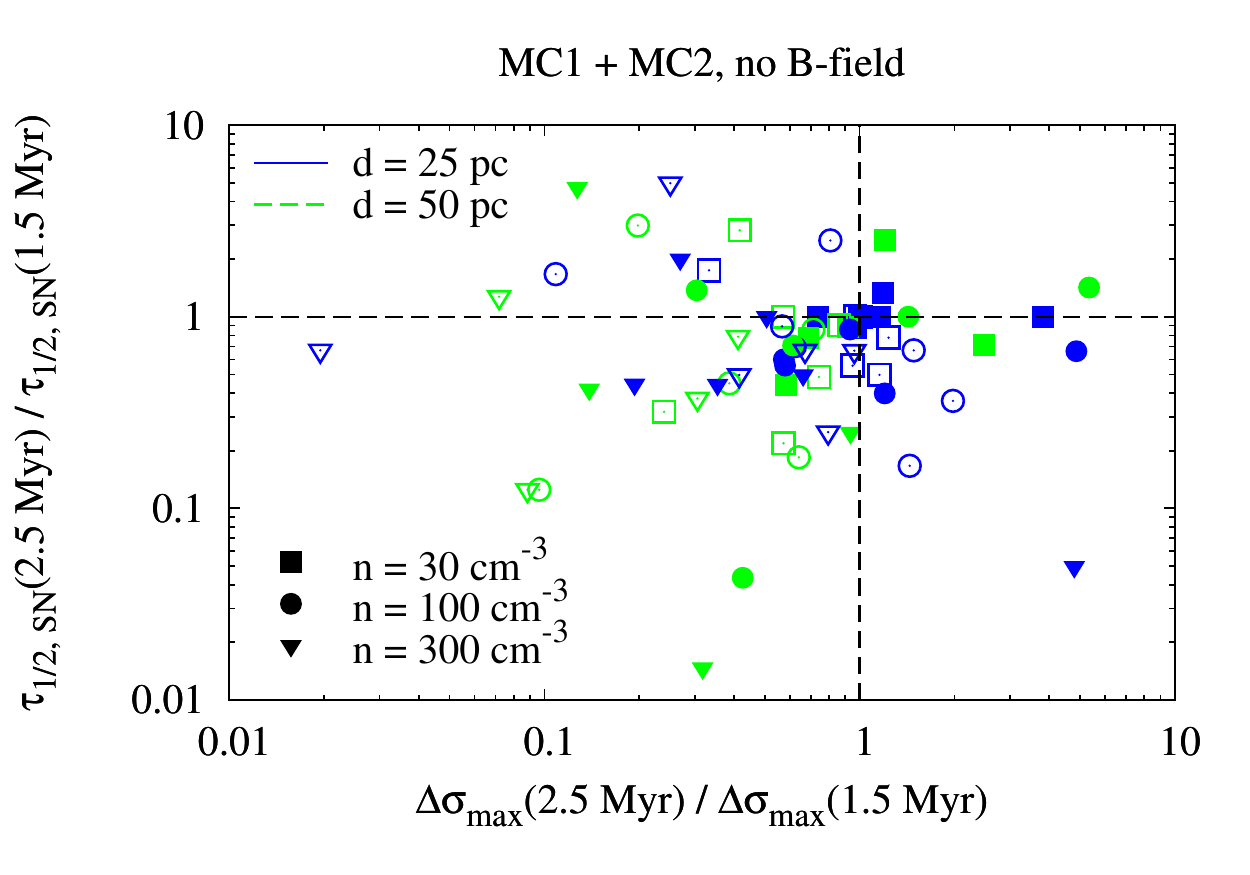}
 \includegraphics[width=0.48\linewidth]{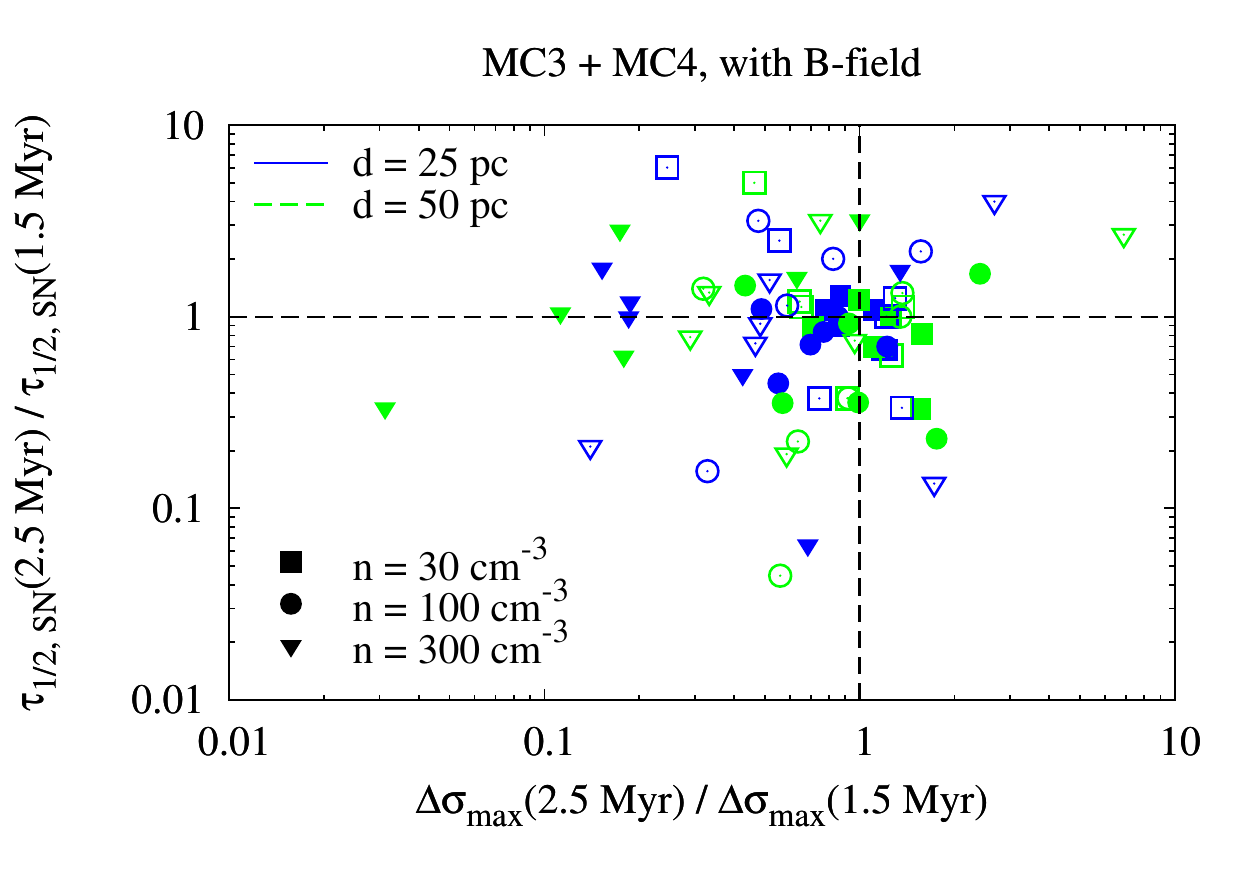}
 \caption{Ratio of the half-life at 2.5 Myr and 1.5 Myr against the corresponding ratio of $\Delta \sigma_\mathrm{max}$ for MC1 (filled symbols) and MC2 (open symbols) without magnetic fields (left panel) and MC3 (filled symbols) and MC4 (open symbols) with magnetic fields (right panel). At the later stage, on average the strength of the SN impact is reduced.}
 \label{fig:sigma_ratio}
\end{figure*}

As can be seen from the mass evolution of MC1 and MC2 (bottom row of Figure~\ref{fig:sigma_mass}), the SNe which go off at $t_\mathrm{SN} = t_0$ + 1.5 Myr impact the clouds during their assembly phase, which is completed around \mbox{$t = t_0$ + 2 Myr}. Similar also applies to the magnetized runs MC3 and MC4 (Figure~\ref{fig:sigma_mass_mag}). In order to test possible differences when a cloud is hit by an SN at a later stage of its evolution, we repeat the previously discussed simulations but now with SNe going off at \mbox{$t_\mathrm{SN} = t_0$ + 2.5 Myr} and \mbox{$t_\mathrm{0, \, mag}$ + 2.5 Myr} for the case without and with magnetic fields, respectively. Since we found that for distances larger than 50 pc the influence is almost negligible (Figures~\ref{fig:sigma_distance} and~\ref{fig:sigma_distance_mag}), we only consider the cases $d$ = 25 and 50 pc.

We repeat the previously described analysis and determine $\Delta \sigma_\mathrm{max}$ and $\tau_\mathrm{1/2, SN}$ (Equations~\ref{eq:sigma_max} and~\ref{eq:tau}). In Figure~\ref{fig:sigma_ratio} we plot the ratio of the half-life times determined for the SNe going off at the two different times (2.5 and 1.5 Myr) against the ratio of $\Delta \sigma_\mathrm{max}$ using the same symbols as in Figure~\ref{fig:decay_time}. The scatter in the ratios seems to be similar for all four clouds considered. Furthermore, we find that for both $d$ the scatter in the distribution is somewhat larger for higher MC density thresholds. We argue that this is a consequence of the shorter dynamical timescales in higher-density regions which allow for more variations in these regions within 1 Myr. For the decay time of the initial gain in velocity, we do not see a clear trend here.

Concerning the strength of the impact, however, SNe going off at later stages seem to have -- on average -- a slightly reduced influence  \mbox{($\Delta \sigma_\mathrm{max}$(2.5 Myr)}/\mbox{$\Delta \sigma_\mathrm{max}$(1.5 Myr) $<$ 1)}. We speculate that this is due to the fact that, as the cloud evolves, its density contrast increases: the already dense regions become even denser and are thus harder to affect with incoming SN shocks. Furthermore, larger low-density regions might evolve due to the accretion of material on the denser regions, which might allow the SN blast to pass relatively unhindered. Finally, at \mbox{$t_\mathrm{SN}$ = $t_0$ + 2.5 Myr} ($t_\mathrm{SN}$ = $t_\mathrm{0, \, mag}$ + 2.5 Myr for the runs with magnetic fields, respectively), also the total masses of the clouds have increased by about 30 -- 40\% (up to a maximum of 5.5 $\times$ 10$^4$ M$_\sun$ for MC1), which also might reduce the influence of an incoming shock carrying a fixed amount of momentum.

\section{Do supernovae drive molecular cloud turbulence?}

In order to assess whether many SNe exploding in the vicinity of an MC can maintain its turbulent state, we estimate the rate of SNe, \mbox{(d SNe/d$t$)$_\mathrm{impact}$}, which could significantly impact the cloud velocity dispersion. Based on our results, we only consider SNe with $d$ $\leq$ 50 pc (Figures~\ref{fig:sigma_distance} and~\ref{fig:sigma_distance_mag}). Applying the Kennicutt-Schmidt relation, we can link the star formation rate to the gas surface density via $\Sigma_\mathrm{SFR} \propto \Sigma_\mathrm{gas}^{1.4}$ \citep[Equation 4 of][]{Kennicutt98}. By assuming a standard IMF \citep{Chabrier01}, which results in 1 SN per 100 M$_{\sun}$ of stars, we then convert $\Sigma_\mathrm{SFR}$ into an SN rate, $\Sigma_\mathrm{SNR}$. This leads to \mbox{15 SNe Myr$^{-1}$ (500 pc)$^{-2}$} for \mbox{$\Sigma_\mathrm{gas}$ = 10 M$_{\sun}$ pc$^{-2}$} \citep[see][]{Walch15,Girichidis16}. With this we obtain the relation
\begin{eqnarray}
\left(\frac{\mathrm{d \, SNe}}{\mathrm{d}t}\right)_\mathrm{impact} &=& \Sigma_\mathrm{SNR} \times \pi \times (50 \mathrm{\, pc)^{2}} \nonumber \\ 
 &\simeq& 0.14 \, \frac{\mathrm{SNe}}{300 \, \mathrm{kyr}} \left(\frac{\Sigma_\mathrm{gas}}{10 \, \mathrm{M}_{\sun} \, \mathrm{pc}^{-2}}\right)^{1.4} \, ,
 \label{eq:snrate}
\end{eqnarray}
where $\left(\frac{\mathrm{d \, SNe}}{\mathrm{d}t}\right)_\mathrm{impact}$ is the number of SNe that hit a cloud in a given time interval under the assumption of random SN locations.

Since we found a typical decay time $\tau_\mathrm{1/2, SN}\simeq$ 300$^{+300}_{-200}$ kyr for the externally driven turbulent component in an MC (see Figure~\ref{fig:decay_time} and right panel of Figure~\ref{fig:sigma_mass_mag}), an MC has to be hit by one SN per 300$^{+300}_{-200}$ kyr to maintain the observed level of MC turbulence of a few 1 km s$^{-1}$ \citep{Larson81,Solomon87,Elmegreen96,Heyer01,Roman10},
which happens for \mbox{$\Sigma_\mathrm{gas}\gtrsim$ 40$^{+49}_{-15}$ M$_{\sun}$ pc$^{-2}$}.
We note that this estimate assumes that SNe in environments with higher $\Sigma_\mathrm{gas}$ are as effective as found in this study, which is not necessarily the case \citep{Li17}.

However, for solar neighbourhood conditions ($\Sigma_\mathrm{gas}$ = 10 M$_{\sun}$ pc$^{-2}$), an MC is hit by a nearby SN once per $\sim$ 2.1 Myr. In the time interval between two hits, the initial gain in $\sigma$ thus decreases to $\sim$ 1\% of its original value. Hence, even for the nearest SNe ($d$ = 25 pc, see Figure~\ref{fig:decay_time}), the increase of the turbulent velocity after \mbox{2.1 Myr} is \mbox{$\lesssim$ 10 m s$^{-1}$}, i.e. significantly lower than the velocity dispersion of a few km s$^{-1}$  found in Galactic MCs with masses comparable to those of the clouds presented here \citep[e.g.][]{Larson81,Solomon87,Roman10}. Therefore, if $\sigma$ were to be maintained by external SNe explosions {\it alone}, this would result in significantly lower values than observed.

On the other hand, in the Galactic Central Molecular Zone (CMZ, typical radius  $\sim$ 250 pc), SN rates of about 1 SN per 1 kyr are reported \citep[e.g.][]{Crocker11,Crocker12,Ponti15}, which translates to $\left(\frac{\mathrm{d \, SNe}}{\mathrm{d}t}\right)_\mathrm{impact}$ $\simeq$ 10 SN per 300 kyr. Based on our study, this rate would be high enough to sustain the MC turbulence by SN explosions. This is in line with recent results by \citet{Kauffmann17}, who found a steep linewidth-size relation in MCs in the Galactic Center, which indicates shock-dominated turbulent motions as e.g. created by SNe.

We argue that -- as we find a comparable effect of SNe for magnetized and unmagnetized clouds (Section~\ref{sec:MHD}) as well as at later times (Section~\ref{sec:laterSN}, here the influence seems to be even slightly reduced) -- the results above should hold for a wide range of evolutionary stages and physical conditions of MCs.

\subsection{Gravity or SN-driven turbulence?}

\begin{figure*}
 \includegraphics[width=\linewidth]{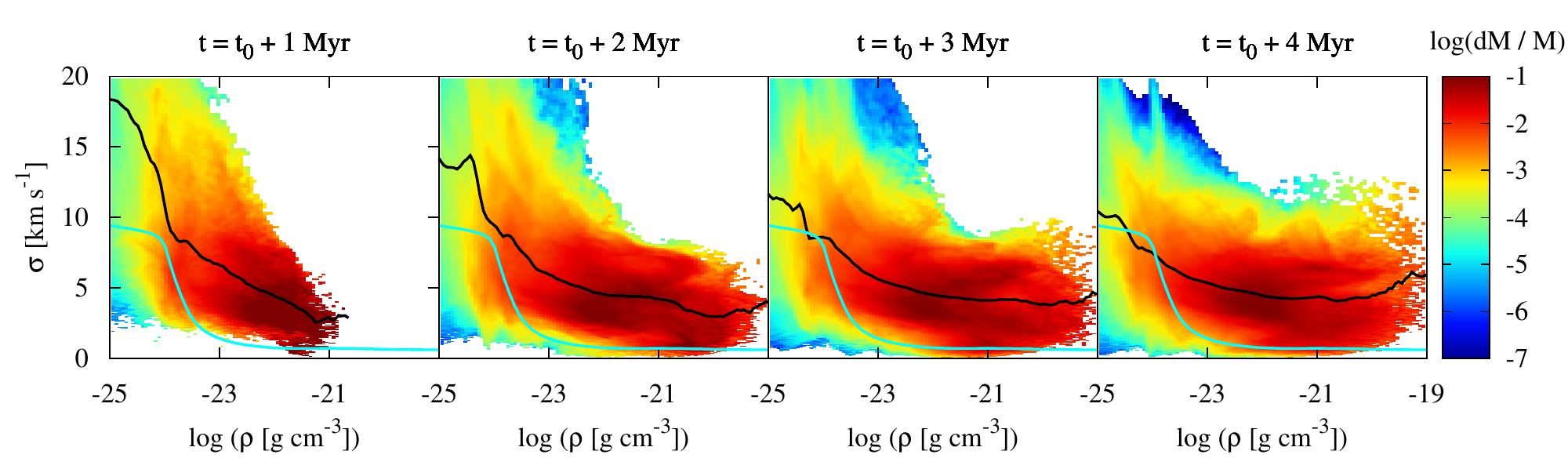}
 \caption{Time evolution of the density-$\sigma$-distribution for MC1. For densities above $\sim$ 10$^{-24}$ g cm$^{-3}$ the mean $\sigma$ in each density bin (black line) is clearly supersonic when compared to the equilibrium sound speed (cyan line) and, at late times, even starts to increase towards the highest densities.}
 \label{fig:sigma_dens}
\end{figure*}

In Figure~\ref{fig:sigma_dens}, we show the gas velocity dispersion in MC1 as a function of density at four different times. Compared to the sound speed, which corresponds to the equilibrium temperature at a given density (cyan line), the mean velocity dispersion in the cloud (black line) is highly supersonic in gas with \mbox{$\rho$ $\gtrsim$ 10$^{-24}$ g cm$^{-3}$}. This could be partly attributed to accretion-driven turbulence \citep{Klessen10,Goldbaum11}. At \mbox{$\rho$ $\gtrsim$ 10$^{-20}$ g cm$^{-3}$}, the velocity dispersion even increases with density, which is a sign for local gravitational collapse.

The latter result is in line with recent simulations by \citet{Ibanez15,Ibanez17} and observations by \citet{Heyer09}, who claimed that gravity is the main driver of turbulence in MCs. Our results also agree with \citet{Iffrig15}, who found that the influence of SNe located outside MCs is only moderate and temporally limited to a few 100 kyr. However, the authors used a highly idealized setup and limit the SN distance to at most 7 pc,  which is why a proper comparison with our work is difficult. 

We emphasize that our results are still in agreement with the scenario in which MCs form from gas that is swept-up by SN explosions \citep{Koyama00,Inoue09,Inutsuka15}. In our simulations, cloud formation is mainly triggered by SNe going off prior to $t_0$, which sweep up turbulent gas while their shocks travel through the three-phase ISM.

However, at first glance our results seem to be in an \textit{apparent} contradiction to \citet{Padoan16}, who found that SNe alone can maintain turbulence in MCs using turbulent box simulations with a comparable resolution of 0.24 pc. They used a rate of \mbox{$\Sigma_\mathrm{SNR}$ = 25 SNe Myr$^{-1}$ (500 pc)$^{-2}$}, which gives $\left(\frac{\mathrm{d \, SNe}}{\mathrm{d}t}\right)_\mathrm{impact}$ = 0.24 SNe per 300 kyr (Equation~\ref{eq:snrate}) and would therefore be too low to drive MC turbulence in a stratified environment. There are, however, significant differences in our setup. \citet{Padoan16} used periodic boundary conditions, which could enhance the impact of SN explosions compared to a stratified disc setup \citep[compare][with \citealt{Girichidis16}]{Gatto15}, as well as a different parameterized cooling function without a chemical network. Furthermore, due to the random distribution of SNe in their simulation domain, a small fraction of SNe might also go off in the clouds themselves, which we do not consider here. Also the different usage of self-gravity can contribute to the differences: whereas \citet{Padoan16} turned on gravity after an initial driving phase of several 10 Myr, in our case gravity is applied from the start of the simulations. In a recent follow-up work, \citet{Padoan17} extended their study to even higher resolutions of 0.0076 pc. However, since the authors only focused on the state of their clouds after self-gravity is switched on, a direct comparison to our results is difficult.

In this context, we note that we have recently investigated the propagation of turbulence from the diffuse into the dense ISM by means of idealized simulations with a resolution of 0.0078 pc.  We find that the transport of turbulence from the low-density environment into the dense medium is inefficient (Rathjen et al. 2018, in preparation), which is in line with our results presented here.

However, we emphasize that, since \citet{Padoan16} did not include self-gravity initially, their clouds might be more diffuse and thus are probably in an earlier evolutionary stage than ours at the time we inject the SNe. As stated before, at this earlier stage, the turbulence of our MCs is mainly generated by SNe, which is in \textit{agreement} with the results of \citet{Padoan16}. Only subsequently, as the clouds become more massive and denser, the importance of external SNe reduces (Section~\ref{sec:laterSN}) and gravity seems to become mainly responsible for maintaining the turbulent stage of MCs. To summarize, our work suggests that -- depending on the evolutionary stage of the cloud -- different turbulence drivers might be at work.

\section{Conclusions}

We present simulations of MC formation within their galactic environment with an effective resolution of \mbox{0.1 pc} including the chemical evolution of the clouds. Our MCs inherit the initial level of turbulence from the diffuse gas they formed in. After they have condensed out of the diffuse ISM, we systematically explore the effect of individual SNe going off outside the clouds at different distances.

Nearby SNe (d $\sim$ 25 pc) boost the turbulent velocity dispersions by up to a few \mbox{1 km s$^{-1}$}, whereas for more distant SNe ($d$ $>$ 50 pc) the boost is almost negligible.
Independent of the distance, however, at these later stages the SNe exploding outside the clouds can increase the level of turbulence only temporarily for a few 100 kyr. Furthermore, we find that the later the SN goes off, the lower is the influence of the SN on the cloud. This holds for MCs with masses from \mbox{$\sim$ 1} to \mbox{5.5 $\times$ 10$^4$ M$_\sun$} and MCs with and without magnetic fields. Furthermore, we tentatively argue that magnetic fields slightly support the transport of turbulence from the diffuse ISM into the dense clouds.

With these results, the following picture for MC formation under solar neighbourhood conditions emerges: in the earliest phase of MC formation, SNe can play a significant role in seeding the initial level of turbulence. At later stages, however, nearby SNe are not able to maintain the turbulent state of MCs. We argue that here turbulence has to be maintained rather by turbulent gas accretion, the onset of local gravitational collapse, and internal stellar feedback. However, in galactic environments with gas surface densities above 40 M$_{\sun}$ pc$^{-2}$, we expect SNe to be able to sustain MC turbulence. In particular in the Galactic CMZ, cloud turbulence driven by SN shocks could be present.

\section*{Acknowledgements}

The authors like to thank the anonymous referee for the comments which helped to significantly improve the paper.
DS and SW acknowledge the support of the Bonn-Cologne Graduate School, which is funded through the German Excellence Initiative. DS and SW also acknowledge funding by the Deutsche Forschungsgemeinschaft (DFG) via the Sonderforschungsbereich SFB 956 \textit{Conditions and Impact of Star Formation} (subproject C5). SW and SH acknowledge support via the ERC starting grant No. 679852 "RADFEEDBACK".
SW, and TN acknowledge support from the DFG Priority Program 1573 ``Physics of the Interstellar Medium''.
PG acknowledges funding from the European Research Council under ERC-CoG grant CRAGSMAN-646955.
TN acknowledges support from the DFG cluster of excellence ''Origin and Structure of the Universe''.
The FLASH code used in this work was partly developed by the Flash Center for Computational Science at the University of Chicago.
The simulations were performed at SuperMUC at the Leibniz-Rechenzentrum Garching.

\cleardoublepage
\appendix
\section{Influence of magnetic fields}

\begin{figure*}
 \includegraphics[width=\linewidth]{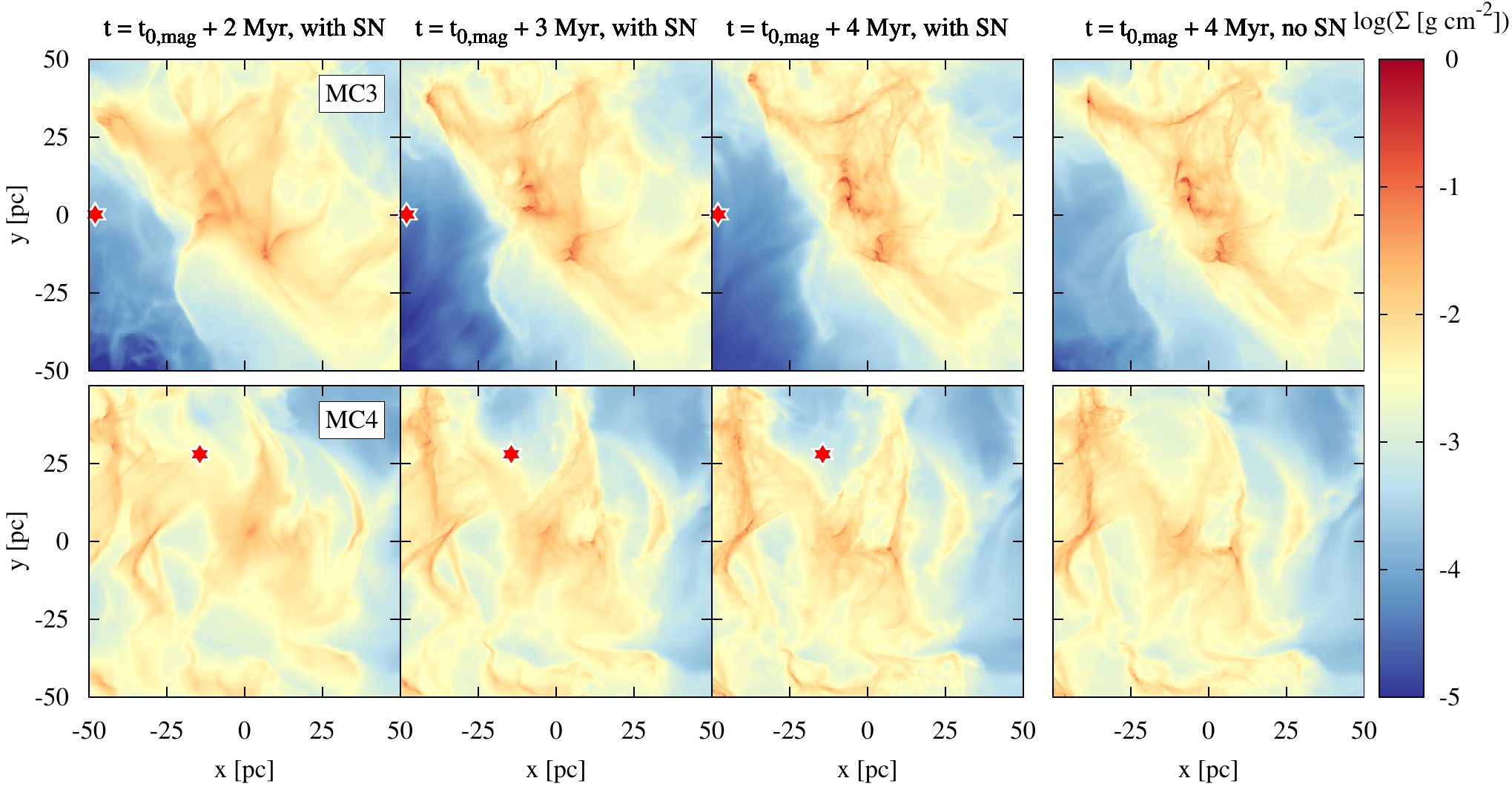}
 \caption{Time evolution of the column density of the runs MC3 and MC4 with magnetic fields exposed to a nearby SN at a distance of 50 pc and 25 pc for MC3 and MC4, respectively. As for unmagnetized runs, mainly the low-density regions are affected. For comparison, the right panels show the reference run without any SN.}
 \label{fig:col_dens_mag}
\end{figure*}

\begin{figure}
\centering
 \includegraphics[width=0.47\linewidth]{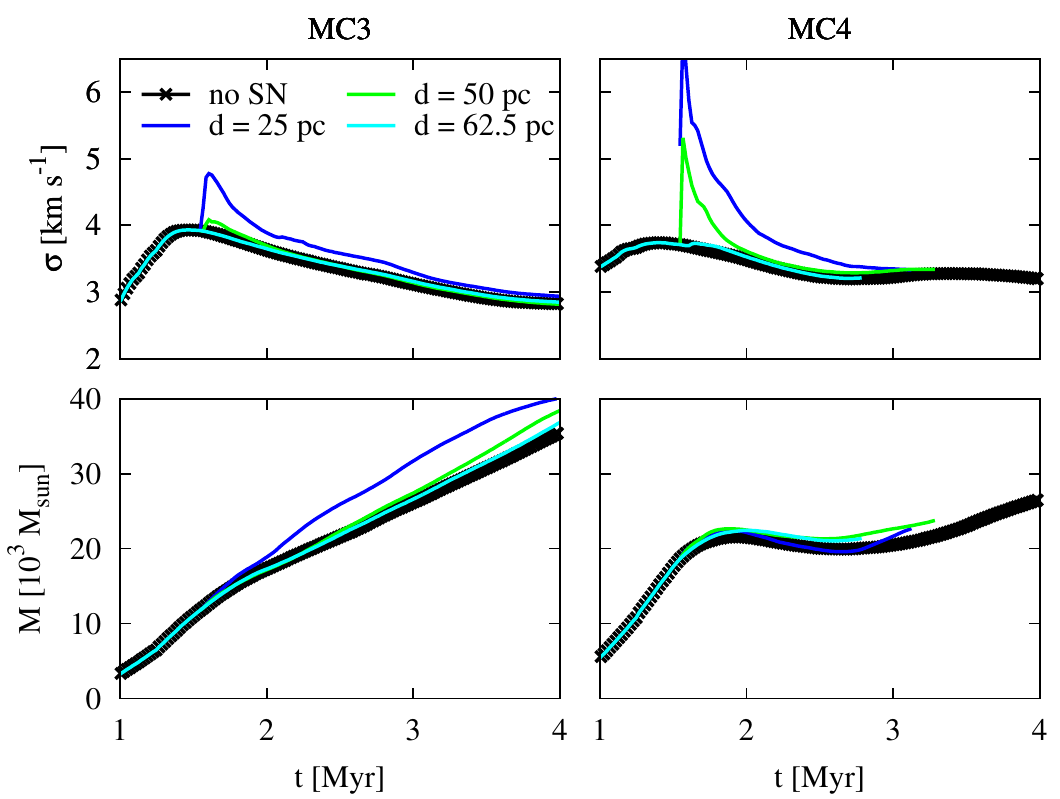}
 \includegraphics[width=0.47\textwidth]{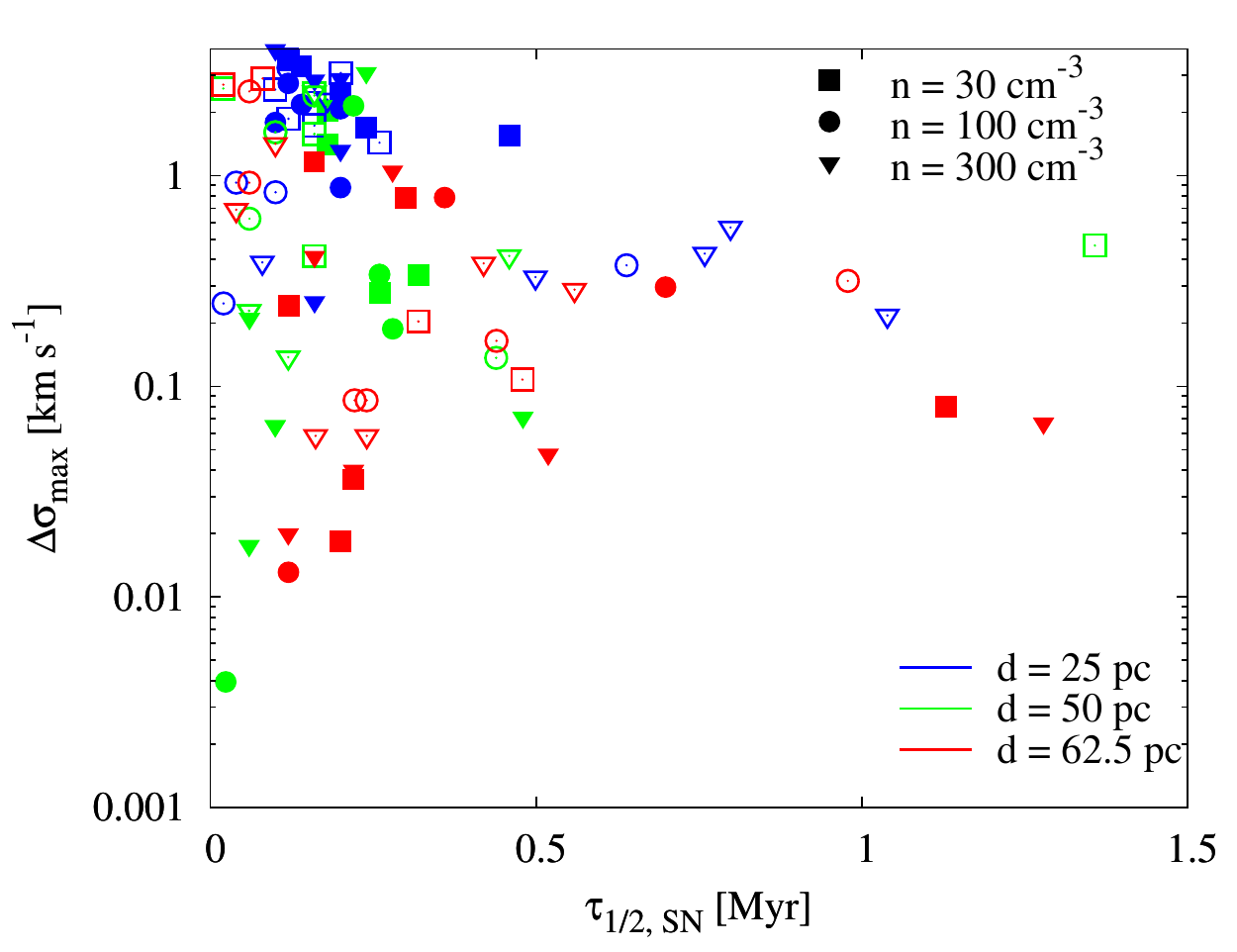}
 \caption{Left: Same as in Figure~\ref{fig:sigma_mass}, but not for the runs MC3 and MC4 with magnetic fields. Right: Same as in Figure~\ref{fig:decay_time}, but for the runs MC3 and MC4. Also for runs including magnetic fields, the effect of SNe seems to be temporally limited to a few 100 kyr. The mass of the clouds is only marginally affected.}
 \label{fig:sigma_mass_mag}
\end{figure}

In Figure~\ref{fig:col_dens_mag} we show the time evolution of the column density of the runs MC3 and MC4 including magnetic fields. Similar to MC1 and MC2, the nearby SNe seem to mainly affect the lower-density regions of the clouds. 

In Figure~\ref{fig:sigma_mass_mag} we show the evolution of the mass and velocity dispersion of gas with \mbox{$n$ $\geq$ 100 cm$^{-3}$}. As for MC1 and MC2, the mass of MC4 remains rather constant at $t$ = $t_\mathrm{0, \, mag}$ + 2 Myr with around 2.5 $\times$ 10$^4$ M$_\sun$, whereas for  MC3 it increases steadily over time to $\sim$ 3.5 $\times$ 10$^4$ M$_\sun$. However, considering the effect of a nearby SN, for both clouds $M_\mathrm{SN}$ shows only a marginal increase of $\lesssim$ 10\%, similar to the runs without magnetic fields. We note that due to computational cost reasons, we did not run all simulations for 4 Myr, but stopped them when the additional velocity gain has dropped to almost zero compared to the reference run.

The velocity dispersion (top row in the left panel of Figure~\ref{fig:sigma_mass_mag}) of the unperturbed clouds (black line with crosses) quickly reaches a roughly constant level of $\sim$ 3 -- 4 km s$^{-1}$ for both MCs. Considering the impact of nearby SNe, we find behaviour very similar to that of the unmagnetized runs: once the SN blast wave impacts the cloud, it experiences an increase which is significantly more pronounced for $d$ $\leq$ 50 pc and subsequently decays within a few 100 kyr. 

This is also reflected in the right panel of Figure~\ref{fig:sigma_mass_mag}, where for each run we show $\Delta \sigma_\mathrm{max}$ and $\tau_\mathrm{1/2, SN}$, following the approach described in Section~\ref{sec:gain}. As for the unmagnetized runs, we find typical values of $\tau_\mathrm{1/2, SN}$ of a few 100 kyr and -- depending on $d$ -- values of $\Delta \sigma_\mathrm{max}$ ranging from $\sim$ 10 m s$^{-1}$ up to a few 1 km s$^{-1}$.

To summarize, overall the runs with magnetic fields thus show behaviors qualitatively and quantitatively similar to those of the runs without magnetic fields.

\end{document}